\documentclass[twocolumn]{aastex62}
\usepackage{nicefrac}

\received{April xx, 2018}
\revised{April xx, 2018}
\usepackage{graphicx}
\usepackage{array}
\usepackage{amssymb}
\usepackage{natbib}
\usepackage{float}
\usepackage{color}
\usepackage[utf8]{inputenc}  
\usepackage{amsmath}  
\usepackage{comment}
\usepackage{footnote}
\usepackage{longtable}
\usepackage[caption2]{ccaption}
\usepackage[figuresright]{rotating}
\usepackage{booktabs}
\usepackage[flushleft]{threeparttable}

\begin{document}
\title{Are Narrow Line Seyfert 1 Galaxies powered by low mass black holes?}
\email{suvenduat@gmail.com}

\author{Gayathri Viswanath}
\affil{Department of Physics and Electronics, CHRIST (Deemed to be University), Bengaluru, Karnataka 560029, India}

\author{C S Stalin}
\affiliation{Indian Institute of Astrophysics, Block II, Koramangala,Bengaluru, Karnataka 560034, India}

\author{Suvendu Rakshit}
\affiliation{Finnish Centre for Astronomy with ESO (FINCA), University of 
Turku, Quantum, Vesilinnantie 5, 20014}

\author{Kshama S. Kurian}
\affiliation{Indian Institute of Astrophysics, Block II, Koramangala,Bengaluru, Karnataka 560034, India}

\author{Ujjwal Krishnan}
\affiliation{Department of Physics and Electronics, CHRIST (Deemed to be University), Bengaluru, Karnataka 560029, India}

\author{Shivappa B. Gudennavar}
\affiliation{Department of Physics and Electronics, CHRIST (Deemed to be University), Bengaluru, Karnataka 560029, India}

\author{Sreeja S. Kartha}
\affiliation{Department of Physics and Electronics, CHRIST (Deemed to be University), Bengaluru, Karnataka 560029, India}


\begin{abstract}
Narrow line Seyfert 1 galaxies (NLS1s) are believed to be powered by accretion of matter onto low mass black holes (BHs) in
spiral host galaxies with BH masses M$_{BH}$ $\sim$ 10$^6$ - 10$^8$ M$_{\odot}$. 
However, the broad band spectral energy distribution  
of the $\gamma$-ray 
emitting NLS1s  are found to be similar to flat spectrum radio quasars.
This challenges our current notion 
of NLS1s having low M$_{BH}$. To resolve this tension 
of low M$_{BH}$ values in NLS1s, we fitted the observed 
optical spectrum of a sample of radio-loud NLS1s (RL-NLS1s), radio-quiet 
NLS1s (RQ-NLS1s) and radio-quiet broad line Seyfert 1 galaxies (RQ-BLS1s) 
of $\sim$500
each   with the 
standard 
Shakura-Sunyaev accretion disk (AD) model.  
For RL-NLS1s
we found a mean log($M_{BH}^{AD}$/M\textsubscript{\(\odot\)}) of 7.98$\pm$0.54.
For RQ-NLS1s and RQ-BLS1s we found mean 
log($M_{BH}^{AD}$/M\textsubscript{\(\odot\)}) of 8.00$\pm$0.43 and 7.90$\pm$0.57,
respectively. While the derived $M_{BH}^{AD}$ values of RQ-BLS1s are
similar to their virial masses, for NLS1s the derived $M_{BH}^{AD}$ 
values are about an order of magnitude larger than their virial estimates. Our
analysis thus indicates that NLS1s have M$_{BH}$ similar to 
RQ-BLS1s and their available  virial 
M$_{BH}$ values are underestimated 
influenced by their observed relatively small emission line widths. 
Considering Eddington ratio as an estimation of the accretion rate and using  
$M_{BH}^{AD}$, we found the 
mean accretion rate of our  RQ-NLS1s, RL-NLS1s and RQ-BLS1s as $0.06^{+0.16}_{-0.05}$, 
 0.05$^{+0.18}_{-0.04}$ and  $0.05^{+0.15}_{-0.04}$, respectively.  Our 
results therefore  suggest that NLS1s have BH masses and
accretion rates similar to  BLS1s. 

\end{abstract}


\section{Introduction} \label{sec:intro}

Narrow Line Seyfert 1 galaxies (NLS1s) are a peculiar class of active galactic 
nuclei (AGN) 
identified 
by \cite{1985ApJ...297..166O}. They are defined to  have
H${\beta}$ emission line full  width at half maximum (FWHM) 
less than 2000 km s$^{-1}$, 
[O III]/H${\beta}$ $<$ 3, strong Fe{\tt II}
emission, steep soft X-ray spectra \citep{1996A&A...309...81W, 
1996A&A...305...53B}, soft X-ray excess \citep{1999ApJS..125..317L,
 1996A&A...305...53B} large amplitude and rapid X-ray variability 
\citep{2017MNRAS.466.3309R}. They
have low mass black holes (M$_{BH}$ $\sim$ 10$^6$ $-$ 10$^8$ M$_{\odot}$) 
and accrete close to the Eddington limit 
\citep{2007ASPC..373..719K,2018MNRAS.480...96W}.
About 5\% of NLS1s emit in the 
radio-band 
\citep{2006ApJS..166..128Z,2017ApJS..229...39R}.
They are more luminous
in the infrared \citep{1996ApJS..106..341M,2007ApJ...654..799R},
less luminous 
in the ultraviolet 
\citep{1997A&A...327...72R,2003PASP..115..592C} and 
show short time scale optical flux variations 
\citep{2004ApJ...609...69K,2013MNRAS.428.2450P,2017MNRAS.466.2679K,
2019MNRAS.483.3036O}. On year like
time scales they show low optical flux variations compared to 
the broad line Seyfert 1 galaxies 
\citep[BLS1s;][]{2017ApJ...842...96R}. They are
found to show flux variations in the infrared bands on long 
\citep{2019MNRAS.483.2362R} and short time scales 
\citep{2012ApJ...759L..31J,2019MNRAS.483.2362R}. 

The urge to understand NLS1s 
increased after the detection of $\gamma$-ray emission in 
about a dozen NLS1s 
\citep{2009ApJ...699..976A,2011nlsg.confE..24F,2012MNRAS.426..317D,
2018arXiv180103963Y,2018ApJ...853L...2P} 
that points to the presence of relativistic jets in them. Detailed
analysis of these $\gamma$-ray emitting
NLS1s ($\gamma$-NLS1s) in the radio band
\citep{2008ApJ...685..801Y,2017A&A...603A.100L} and
the broad band spectral energy distribution modelling 
\citep{2014ApJ...789..143P,2016ApJ...819..121P, 2018ApJ...853L...2P}
indicate that they have many properties similar to the blazar (flat spectrum
radio quasar, FSRQ) category 
of AGN. The two properties in which NLS1s differ from 
blazars  are that NLS1s have low mass black holes  (BHs)
in spiral hosts, whereas blazars are powered by high mass BHs in 
elliptical hosts. 

From spectro-polarimetric observation of
a $\gamma$-NLS1 PKS 2004$-$447, 
\cite{2016MNRAS.458L..69B} found a $M_{BH}$ of  
6 $\times 10^8$ M\textsubscript{\(\odot\)}, 
larger than the value of 5 $\times 10^6$ M\textsubscript{\(\odot\)} from 
the total intensity spectrum. 
\cite{2013MNRAS.431..210C}, by fitting accretion
disk (AD) models to the spectra of 23 radio-loud NLS1s (RL-NLS1s),  
found them to
have $M_{BH}$ similar to blazars. 
Also  
it was shown that fitting AD  model to 
type-1 AGN spectra gives realistic
estimates of $M_{BH}$ \citep{2015MNRAS.446.3427C,2016MNRAS.460..212C} compared to virial 
estimates, as virial estimates
are prone to uncertainties \citep{2018NatAs...2...63M}. 
Thus, available studies point to drawbacks in virial BH 
mass estimates, and therefore the current notion that NLS1s
have low mass BHs in them from virial estimates  needs a critical 
evaluation. 
Considering the current belief that powerful jets can only be fueled 
by elliptical galaxies with BH masses of the order of $10^8$ to 
$10^9$ M\textsubscript{\(\odot\)}, the so called 
{\it``elliptical-jet paradigm"} \citep{2011nlsg.confE..24F}, NLS1s have 
become important candidates to test this hypothesis, particularly after the 
detection
of $\gamma$-rays in a handful of sources.  The motivation for this 
work is to check if RL-NLS1s are indeed
powered by low mass BHs. 

\section{Sample} \label{sec:sample}
\subsection{Radio-loud NLS1s}
Our  sources are from the catalog of   NLS1s 
by \cite{2017ApJS..229...39R}. 
We cross-correlated the 11,101 NLS1s with the Faint Images
of the Radio Sky at Twenty cm (FIRST) survey within a search radius of 
2 arcsec 
to find radio-emitting NLS1s. 
This led us to a sample
of 554 NLS1s detected in FIRST, which is around 5\% of the total
sample of NLS1s. For this work we considered the sources
that are detected in FIRST as radio loud and the sources not detected in FIRST as radio-quiet. Our sample has an average radio loudness of log R = 
1.32, where R = F$_{(5GHz)}$/F$_{(B\mathrm{band})}$, with F$_{(5GHz)}$ and 
F$_{(B\mathrm{band})}$
being the flux densities in the radio band at 5 GHz and optical B-band, respectively. 
\subsection{Radio-quiet NLS1s and BLS1s}
To check for any differences in the derived BH masses of RL-NLS1s, 
relative to radio-quiet NLS1s (RQ-NLS1s) and BLS1s, we also 
selected a  control sample of RQ-NLS1s and  radio-quiet BLS1s (RQ-BLS1).
For each RL-NLS1, we selected a RQ-NLS1 and a RQ-BLS1 
matched in redshift and optical g-band brightness.  Thus as a control
sample, we selected 554  RQ-NLS1s and 471 RQ-BLS1s. 

\section{Analysis} \label{sec:analysis}
We derived BH masses for all our sample of 
RL-NLS1s, RQ-NLS1s and RQ-BLS1s  using
two procedures (a) virial method (VM) and (b) fitting
AD model to the observed SDSS spectra.  While the BH masses for RL-NLS1s and RQ-NLS1s
using VM method were taken from \cite{2017ApJS..229...39R}, for RQ-BLS1s, they were
estimated using the procedures in Section 3.1
\subsection{M$_{BH}$ using virial method}
The 
width of the broad emission lines in AGN spectra can serve as a proxy for the 
velocity of the clouds in their broad line region (BLR)
and in virial equilibrium, the mass of the BH is related
to the observed width of the emission lines as
\begin{equation}
M_{BH}^{VM} = \frac{f R_{BLR} \Delta V^2}{G}
\end{equation}
where, G is the gravitational constant, R$_{BLR}$ is the average radius of the 
BLR from the central black hole, $\Delta$ V is the FWHM of the emission line
and $f$ is the scale factor that accounts for the geometry of the BLR. 
We determined R$_{BLR}$ using the following scaling relation
between  the monochromatic luminosity at 5100 \AA ~and R$_{BLR}$ 
\begin{equation}
log\left(\frac{R_{BLR}}{\mathrm{1 ~light ~day}}\right) = A + B \times log\left(\frac{\lambda L_{\lambda}(5100 \AA)\mathrm{~erg \, sec^{-1}}}{10^{44}}\right)
\end{equation}
Here, the values of A and B were taken from \cite{2013ApJ...767..149B}. Taking the 
value of $\lambda L_{\lambda}$ and the FHWM of the H${\beta}$ emission line
and adopting $f$ = 3/4 \citep{2017ApJS..229...39R}, we derived 
virial BH masses using
Equations 1 and 2. For RL-NLS1s, the mean virial BH mass
is  log($M_{BH}^{VM}$/M\textsubscript{\(\odot\)}) = 6.98$\pm$0.49. For the sample of
RQ-NLS1s and RQ-BLS1s, we found mean
values of log($M_{BH}^{VM}$/M\textsubscript{\(\odot\)}) = 7.07$\pm$0.38
and 8.01$\pm$0.48 respectively. Thus, based on VM,  RQ-BLS1s have
larger mass BHs than NLS1s.

\begin{figure}
\vbox{
\hspace*{-2.2cm}\includegraphics[scale=0.28]{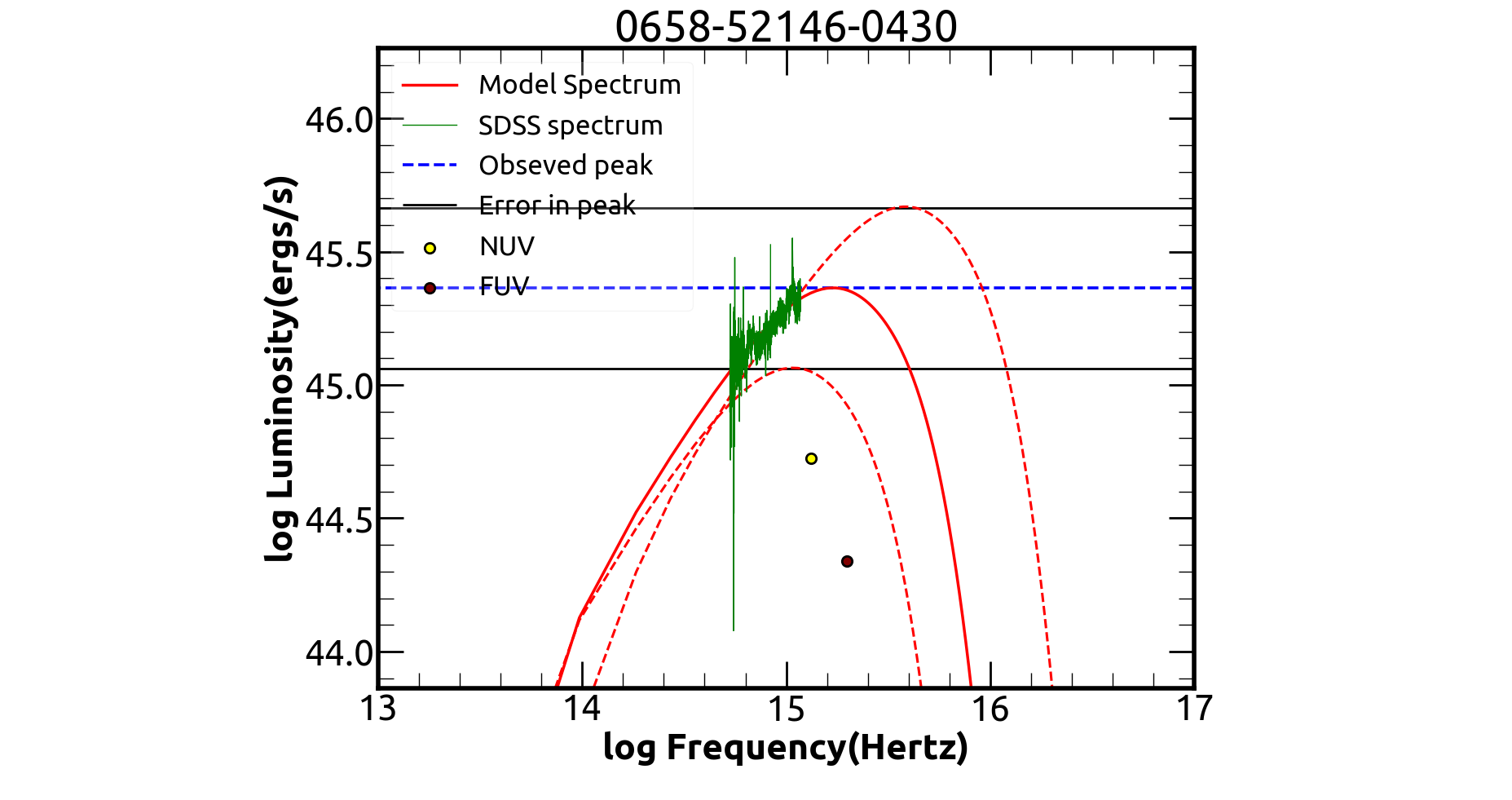}
\hspace*{-2.2cm}\includegraphics[scale=0.28]{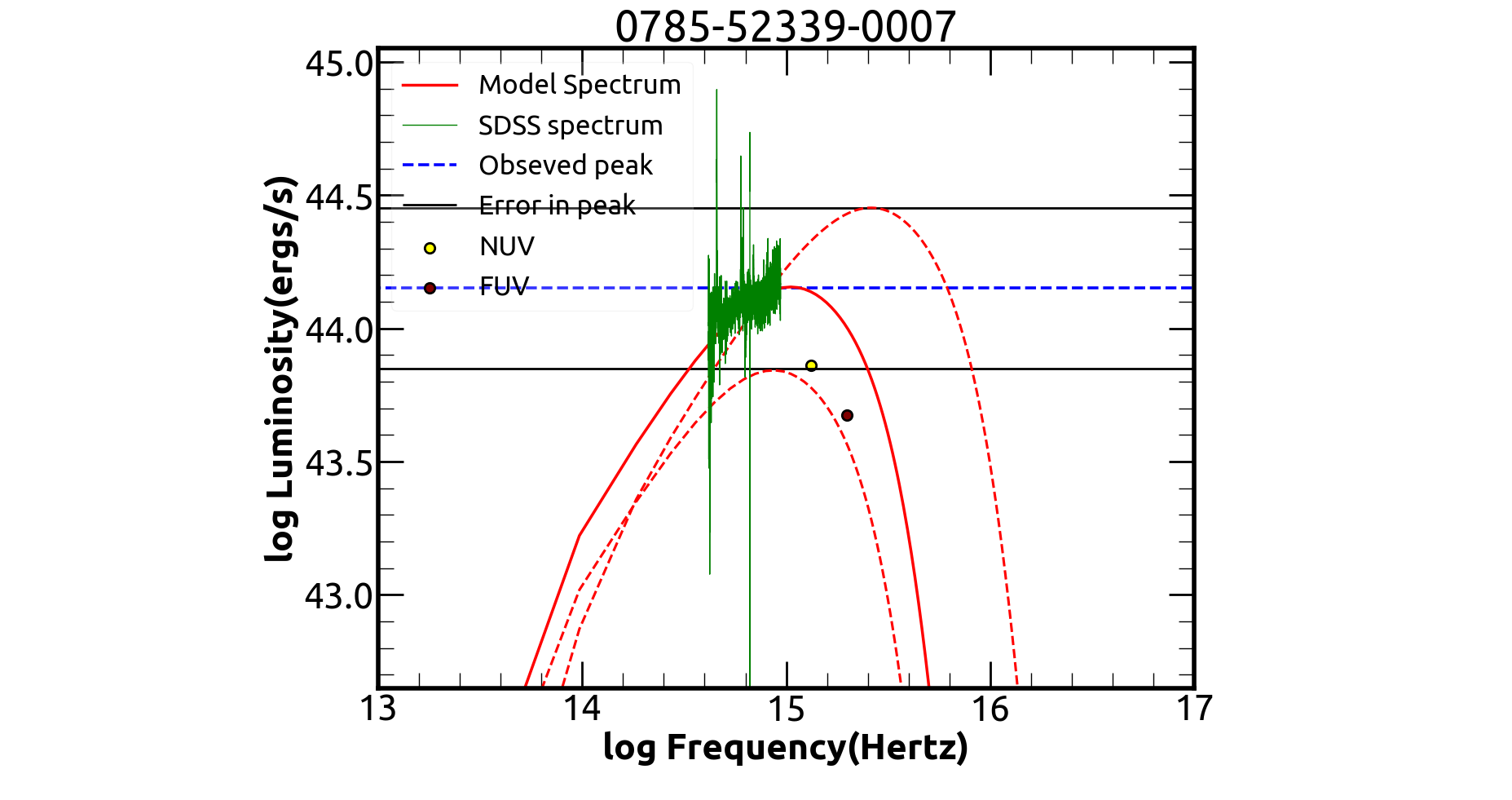}
}
\caption{AD fits to the observed spectrum for two sources: 
0658-52146-0430 (top panel),  and
0785-52339-0007 (bottom panel). Here green is the observed SDSS spectrum, 
the red solid line is the calculated AD spectrum. The blue dashed line is 
the peak luminosity from the AD of the source derived from the H$\beta$ 
luminosity. The solid black lines are the 1 sigma error in the peak 
luminosity, that were
used to estimate the error in M$_{BH}^{AD}$. The filled circles are the GALEX measurements. They were not included in the fitting process. 
\label{Fits}}
\end{figure}

\subsection{M$_{BH}$ using AD model fitting}
Fitting Shakura-Sunyaev  (S\&S) AD model \citep{1973A&A....24..337S}  to estimate
M$_{BH}$ is known and has been applied to 
type 1 AGN \citep{2015MNRAS.446.3427C,2016MNRAS.460..212C} and
RL-NLS1s \citep{2013MNRAS.431..210C}. This technique is better
than the virial M$_{BH}$ method that is  affected
by uncertainties \citep{2018NatAs...2...63M} like (a) incomplete 
knowledge on the distribution of gas clouds (b) inclination of the AD to 
the line of sight  and (c) 
dependence of the scale factor on inclination. For this work, we followed the 
procedure in \cite{2013MNRAS.431..210C} and described in brief below. 
We assumed a simple, non-relativistic, geometrically thin, optically thick 
AD in steady state, whose thermal emission is described by a 
standard S\&S AD model.  
Each annulus of the AD emits
black body radiation at a temperature, which is a function of radius R
of the disk T(R), and the emitted spectrum is a superposition of several 
black body spectra. For evaluating the emitted spectrum 
the  inner radius of the BH ($R_{in}$) was taken as 6$R_g$ and the outer radius 
as 2000$R_g$, where $R_g$ = G$M_{BH}$/$c^2$ is the gravitational radius of 
the BH. The radiative efficiency is $\eta$ = 
$R_g$/2$R_{in}$ and $M_{BH}^{AD}$ depends on both $R_{in}$ and $\eta$. The 
de-projection factor for calculating the isotropic disk luminosity was taken as 
$< 2\cos \theta > =1.7$ corresponding to an average 
viewing angle ($\theta$) of $30^{\circ}$. The AD model created 
using the above assumptions varies with $M_{BH}^{AD}$, mass accretion rate 
$\dot{M}$ and $\eta$. The peak frequency ($\nu_p$) of 
the generated AD spectra and the luminosity at the 
peak frequency ($L_{\nu_p}$) scales  as 
\begin{equation}
\nu_p \propto \Big(\frac{\eta}{0.1}\Big)^{\nicefrac{3}{4}} \times \Big(\frac{M_{BH}^{AD}}{10^6 M\textsubscript{\(\odot\)}}\Big)^{\nicefrac{-1}{2}} \times \Big(\frac{\dot{M}}{M\textsubscript{\(\odot\)} yr^{-1}}\Big)^{\nicefrac{1}{4}}
\end{equation}

\begin{equation}
\nu_p {L_{\nu}}_p \propto \Big(\frac{\eta}{0.1}\Big) \times \Big(\frac{\dot{M}}{M\textsubscript{\(\odot\)} yr^{-1}}\Big)  
\end{equation}

For any given value of $\eta$, from an estimation of luminosity and 
$\nu_p$, the parameters $M_{BH}^{AD}$ and $\dot{M}$ can be determined.
We obtained the isotropic disk luminosity 
using the sum of the broad and narrow component fluxes of $H\beta$ 
as:
\begin{equation}
L^{iso}_d (BLS1)  = 303 \times L(H\beta)
\end{equation}

\begin{equation}
L^{iso}_d (NLS1)  = 424 \times L(H\beta)
\end{equation}

where $L(H\beta)$ is the luminosity of the $H\beta$ line
calculated from the derived fluxes using the procedures given in \cite{2017ApJS..229...39R}. 
From the  isotropic disk luminosity, we calculated the peak luminosity from the AD model as:

\begin{equation}
\nu_p L_{\nu_p} = 0.5 \times L^{iso}_d
\end{equation}
This fixes a 'ceiling' to the theoretical AD spectrum that was 
created. The error in H$\beta$ fluxes and the uncertainty of 
$_{\widetilde{~}}$2 in Equations 5 and 6 
\citep{2013MNRAS.431..210C}, 
were propagated in the calculations to obtain an 
upper and lower limit to the peak disk luminosity $\nu_p L_{\nu_p}$.

The optical spectra for AD modeling
were from 
SDSS DR-12. To avoid  
instrumental noise, we dropped  
bins equivalent to cover 110 \AA from both the ends of each spectrum. 
The spectra were de-reddened 
following \cite{1989ApJ...345..245C} and  brought to the 
rest frame. The contribution of the 
host galaxy to the observed spectra was removed following the procedure
in \cite{2017ApJS..229...39R}. Each of the resultant 
(de-reddened and host galaxy subtracted) spectrum
was then matched with the generated theoretical AD spectrum.

The theoretical AD spectrum was first generated assuming an initial $M_{BH}^{AD} = 
5 \times 10^4 M\textsubscript{\(\odot\)}$, $\eta = 0.1$ (fixed) and $\dot{M} = 
1.0 M\textsubscript{\(\odot\)} yr^{-1}$ .
The value of $\dot{M}$ was iteratively increased or 
decreased till the peak of the AD spectrum lies on the
line defining the observed peak luminosity (shown as a blue dotted line in Figure \ref{Fits})  from the source. 
The value of $\dot{M}$ that gave the minimum $\chi^2$ between the peak of the AD spectrum
and the line luminosity obtained via a fit of $\chi^2$ against $\dot{M}$ was considered as 
the final $\dot{M}$.
Once this was 
achieved, $M_{BH}^{AD}$  was increased in steps of 
5 $\times$ $10^4$ M\textsubscript{\(\odot\)}. This shifts the theoretical
AD spectrum horizontally. 
We chose two anchor points one around 2900 \AA 
~and the other around 3500 \AA ~and evaluated $\chi^2$ at those two anchor points between the theoretical
AD spectrum and the SDSS spectrum. This iteration was continued till we attained a minimum $\chi^2$ 
through a fit of $\chi^2$ against $M_{BH}^{AD}$.
This constrains the BH mass 
of the source. The fitting was repeated for the upper and lower 
error limits of the peak disk luminosity (indicated by the solid black lines in 
Figure \ref{Fits}), to find the confidence limits in the estimated value
of $M_{BH}^{AD}$. 

\begin{figure}
        \hspace*{-0.6cm}\includegraphics[scale=0.2]{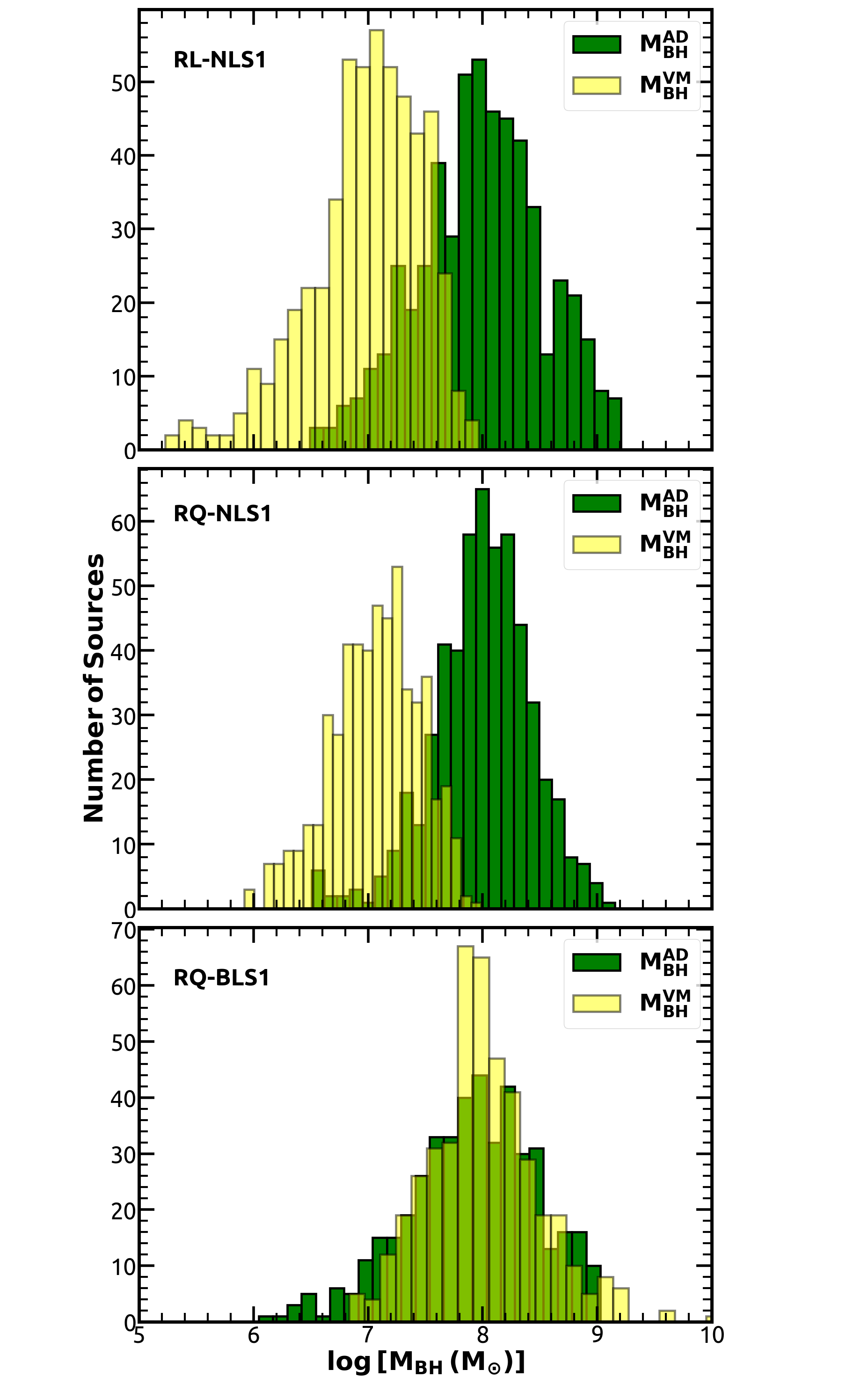}
        \caption{Distribution of logarithmic of M$_{BH}$ in units of M$_{\odot}$ for different
categories of sources in our sample. 
The yellow and green histograms are for  M$_{BH}^{VM}$ and $M_{BH}^{AD}$ respectively. 
\label{Histo}}
\end{figure}

\section{Results}
\subsection{M$_{BH}$ of RL-NLS1s, RQ-NLS1s and BLS1s}
The AD fitting  was carried out on RL-NLS1s, 
RQ-NLS1s and RQ-BLS1s each consisting of 554 sources, except RQ-BLS1s 
that contain 471 sources. Of these, our automatic fitting procedure
converged for 537 RL and RQ-NLS1s and 448 RQ-BLS1s.
Spectral fits to two RL-NLS1s from  our sample are
shown in Figure \ref{Fits} and the results  
are given in Table \ref{Fit_table}. In the same table
are given M$_{BH}^{VM}$ obtained for RQ-BLS1s  using the VM 
 outlined in 
Section 3.1 and taken from \cite{2017ApJS..229...39R} for NLS1s. 
Also given are the accretion rate ($\lambda_{Edd}$)  calculated as 
$\lambda_{Edd}$ = $424 \times L(H\beta)/L_{Edd}$  
for NLS1s and $\lambda_{Edd}$ = $303 \times L(H\beta)/L_{Edd}$ for BLS1s,
where $L_{Edd}$ is the Eddington luminosity
defined as $L_{Edd} = 1.3 \times 10^{38} \left(M_{BH}^{AD}/M_{\odot}\right)$ 
erg s$^{-1}$. For 
RQ-NLS1s, RL-NLS1s and RQ-BLS1s the calculated mean values of 
$\lambda_{Edd}$ are $0.06^{+0.16}_{-0.05}$, $0.05^{+0.18}_{-0.04}$ and $0.05^{+0.15}_{-0.04}$,
respectively. 

\begin{deluxetable*}{ccccccccc}
\tablecaption{Results of AD model fitting to the  SDSS spectra.  
The table in full is available in the electronic version of the article \label{Fit_table}.}
\tablewidth{0pt}
\tablehead{\colhead{$\alpha_{2000}$ } & \colhead{$\delta_{2000}$ } & \colhead{$z$} &  \colhead{Type} & \colhead{R} & 
\colhead{log L(H$_\beta$)} & \colhead{log($M_{BH}^{VM}/M_{\odot}$)} &  \colhead{log($M_{BH}^{AD}/M_{\odot}$)} & \colhead{$\lambda_{Edd}$}  \\ 
\colhead{} & \colhead{}  & \colhead{} & \colhead{} & \colhead{} & \colhead{(ergs/s)} & \colhead{} & \colhead{} & \colhead{} } 
\startdata
 00:09:39.82 &	 +13:27:17.0 & 0.482 & RQ-NLS1	 & ----	  & 41.75	 & 6.82   &	7.57$^{+0.61}_{-0.39}$  & 0.05$^{+0.07}_{-0.04}$ \\
 00:08:04.17 &  -01:29:17.0 & 0.314 & RL-NLS1  & 1.848  & 41.80   & 6.97   & 	7.29$^{+0.44}_{-0.35}$  &  0.11$^{+0.13}_{-0.07}$ \\
 00:11:37.25 &  +14:42:01.4 & 0.132 & RL-NLS1  & 0.711  & 41.93   & 7.16   &	7.69$^{+0.47}_{-0.37}$  &  0.06$^{+0.07}_{-0.04}$ \\
\enddata
\end{deluxetable*}

\begin{figure}
\vbox{
\hspace*{-0.6cm}\includegraphics[scale=0.2]{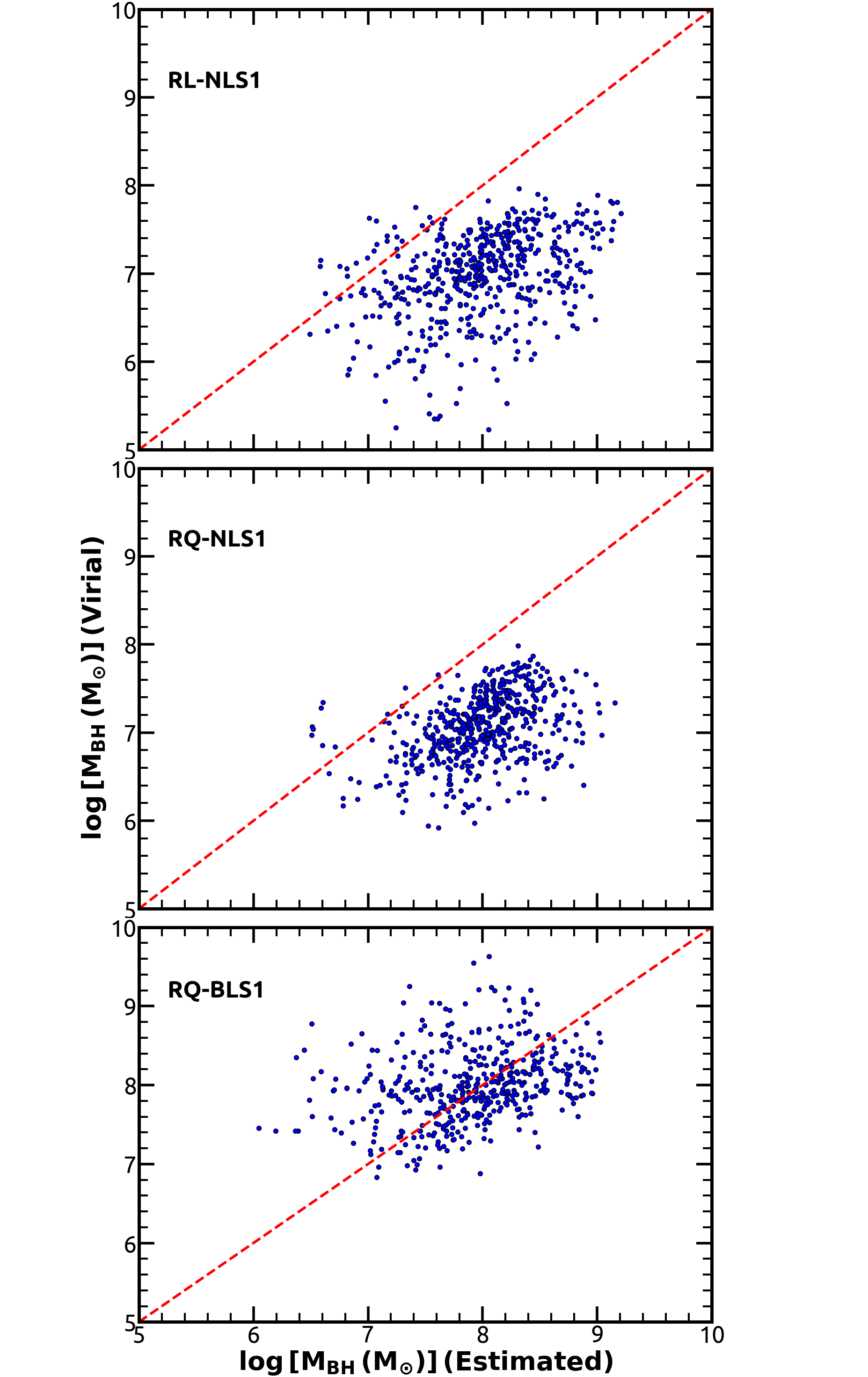}
    }
\caption{A comparison of the estimated M$_{BH}^{VM}$ and 
M$_{BH}^{AD}$ values for different categories of AGN. The red dashed line is for 
M$_{BH}^{VM}$ = M$_{BH}^{AD}$.
\label{Histo-1}}
\end{figure}

In Figure \ref{Histo} (top panel) we show the  distribution of  
$M_{BH}^{AD}$ and $M_{BH}^{VM}$  for RL-NLS1s.  
The mean value of 
log\Big($\frac{M_{BH}^{AD}}{M\textsubscript{\(\odot\)}}$\Big) 
is 7.98 $\pm$ 0.54. This is larger than the mean  
log\Big($\frac{M_{BH}^{VM}}{M\textsubscript{\(\odot\)}}$\Big) of
6.98 $\pm$ 0.49. A two sample Kolmogorov-Smirnov (KS) test 
at a significance of $\alpha$ = 5\% confirms that the two distributions are  
different with a test statistics (D) of 0.70 and a null-hypothesis (the two distributions are identical) probability  $p$ of 6.4 $\times$ 10$^{-116}$ . 
The distributions of 
$M_{BH}^{VM}$ and $M_{BH}^{AD}$ for RQ-NLS1s are shown in 
the middle panel of Figure \ref{Histo}. The two distributions are different
with mean values of 7.07$\pm$0.38 and 8.00$\pm$0.43
for  $M_{BH}^{VM}$ and $M_{BH}^{AD}$  respectively. This is confirmed by the KS test with D = 0.78 and p = 2.6 $\times$ 10$^{-146}$. The bottom panel of
Figure \ref{Histo} shows the distributions of $M_{BH}^{AD}$ and $M_{BH}^{VM}$
obtained for RQ-BLS1s. The distributions are
nearly identical with mean values of 7.90$\pm$0.57 and 8.01$\pm$0.48 for
$M_{BH}^{AD}$ and $M_{BH}^{VM}$  respectively. Though KS test rejects the null hypothesis with p=0.01, D has a small value of 0.11. Our analysis 
thus indicates that in both RL-NLS1s and 
RQ-NSL1s, $M_{BH}^{AD}$ values are  
systematically larger than $M_{BH}^{VM}$. However, in the case 
of RQ-BLS1s, both the estimates are not systematically different.
This is evident from the plots in Figure \ref{Histo-1}. In the
$M_{BH}^{AD}$ v/s $M_{BH}^{VM}$ diagram, for both RL-NLS1s and
RQ-NLS1s the points are systematically away from the 
$M_{BH}^{AD}$ = $M_{BH}^{VM}$ line. In the  case of RQ-BLS1s
the points are scattered around the line, with the mean value
of log\Big($M_{BH}^{AD}$/$M_{BH}^{VM}$\Big) =  -0.11$\pm$0.64. For RL-NLS1s and 
RQ-NLS1s, we found mean log\Big($M_{BH}^{AD}$/$M_{BH}^{VM}$\Big) values
of 1.00$\pm$0.57 and 0.93$\pm$0.45 respectively.

\subsection{M$_{BH}$ of $\gamma$-NLS1s}
A total of 16  NLS1s are found to be emitters of 
$\gamma$-rays \citep{2019ApJ...872..169P}. 
Of these, we have 9  $\gamma$-NLS1s in our sample. The 
values of M$_{BH}^{AD}$ obtained for these
 9 sources are given in Table \ref{table-2}. For these sources  M$_{BH}^{AD}$
values are larger that $M_{BH}^{VM}$ except one. 

\begin{deluxetable}{cccc}
\tablecaption{Black hole masses of $\gamma$-NLS1s 
in our sample} 
\tablewidth{0pt}
\tablehead{\colhead{$\alpha_{2000}$} & \colhead{$\delta_{2000}$} & \colhead{M$_{BH}^{AD}$} & \colhead{M$_{BH}^{VM}$} \label{table-2}} 
\startdata
08:49:57.98 & +51:08:29.1 &	7.86$^{+0.64}_{-0.40}$ & 7.38 \\
09:32:41.15 & +53:06:33.8 &	8.01$^{+0.53}_{-0.38}$ & 7.45 \\
09:48:57.33 & +00:22:25.5 &	8.97$^{+0.01}_{-0.69}$ & 7.30 \\
12:46:34.65 & +02:38:09.1 &	8.63$^{+0.27}_{-0.54}$ & 7.21 \\
14:21:06.04 & +38:55:22.8 &	8.63$^{+0.30}_{-0.52}$ & 7.36 \\
15:20:39.69 & +42:11:11.2 &	7.07$^{+0.35}_{-0.27}$ & 7.60 \\
16:44:42.53 & +26:19:13.3 &	8.30$^{+0.46}_{-0.47}$ & 6.98 \\
21:18:17.40 & +00:13:16.8 &	7.98$^{+0.78}_{-0.42}$ & 7.25 \\
21:18:52.97 & -07:32:27.6 &	7.94$^{+0.78}_{-0.40}$ & 6.98 \\ \hline
\enddata
\end{deluxetable}

\section{Discussion}

It is likely that M$_{BH}^{AD}$  values are close
to the true  BH masses in AGN, as this
technique depends only on the ability to match the theoretical
AD spectra to the observed SDSS spectra and is independent of 
the geometry and kinematics of the BLR \citep{2018NatAs...2...63M}. 
The limitation here is the wavelength coverage of the SDSS spectra. 
Increased wavelength coverage into the UV region 
using data from GALEX could be an advantage, however we have not attempted here. 
This is because
the SDSS spectra and the GALEX observations pertain to different epochs
and our sources would have varied during those two epochs. 
Another important factor that can affect the M$_{BH}^{AD}$ 
values is related to the contribution of relativistic jets
to the SDSS spectra. This uncertainty will be 
there in the case of RL-NLS1s, however, unlikely to be present
in RQ-NLS1s and RQ-BLS1s. 
We did not
attempt to correct for this effect (see \citealt{2013MNRAS.431..210C}), firstly, because of the non-simultaneity of the 
infra-red measurements and SDSS spectra and secondly on the possibility of the sources
in a faint activity state during the epoch when the SDSS spectra were taken 
leading to low/no contribution of jet emission to the spectra.  
Though AGN flux variability properties
can in principle have some effect on AD model fits, 
they are unlikely to have any systematic effects on the estimated M$_{BH}^{AD}$ values.

Though AD model fits to SDSS spectra to derive BH masses have the limitations described
above,  M$_{BH}^{VM}$ estimation method too  suffer from
uncertainties like (i) lack of our knowledge on the geometry and 
kinematics of BLR and (ii) inclination of the source relative to the observer.
From the $M_{BH}^{AD}$ values obtained for 
NLS1s, it is clear that our earlier knowledge of BH masses in them based on virial estimates is an underestimation.
For our sample of 537 RL-NLS1s (that also
includes 9 $\gamma$-NLS1s) we found
mean log($M_{BH}^{AD}/M_{\odot}$) of 7.98$\pm$0.54. For our sample of 
RQ-NLS1s and RQ-BLS1s we found mean log($M_{BH}^{AD}/M_{\odot}$)
values of 8.00$\pm$0.43 and 7.90$\pm$0.57 respectively. Thus our AD model fits to 
all the three categories of sources in a homogeneous manner point to similar
BH masses in all the three categories. 
This leads us to conclude that NLS1s are not powered by low mass BHs, 
instead have BH masses similar to RQ-BLS1s and blazars. 
Report for large BH masses in NLS1s are available in 
literature from  
 AD model fits \citep{2013MNRAS.431..210C,2018rnls.confE..44C} and 
spectro-polarimetry \citep{2016MNRAS.458L..69B}. 
Focussing only on the sub-set of 9
$\gamma$- NLS1s in our sample, 
we found mean log($M_{BH}^{AD}/M_{\odot}$) of 8.15$\pm$0.56. 
We are therefore inclined to argue that $\gamma$-NLS1s  can 
no-longer be considered the {\it ``low mass BH counterparts to FSRQs"}.

An explanation for the narrow width of broad emission
lines in NLS1s and subsequently an underestimation of $M_{BH}^{VM}$
in them could be due to the assumption of these
sources having a disk like BLR and viewed face on \citep{2008MNRAS.386L..15D}.
To probe
the effects of viewing angle on the $M_{BH}^{AD}$ from
AD fitting, we derived BH masses for the RL-NLS1s assuming a 
viewing angle of $\theta$ =  $5^{\circ}$, which is typical of
$\gamma$-ray emitting AGN. For RL-NLS1s we obtained
mean log($M_{BH}^{AD}/M_{\odot}$) of 7.94 $\pm$ 0.54 which is similar to the mean 
log($M_{BH}^{AD}/M_{\odot}$) of 7.98 $\pm$ 0.54 obtained for the same sample considering
a viewing angle of $30^{\circ}$. Therefore, AD model fits to the observed spectrum to
find BH masses is less dependent on the viewing angle (see also \citealt{2018NatAs...2...63M}). Also, \cite{2008ApJ...678..693M} has
proposed that the BH masses of NLS1s determined from optical 
spectroscopy can be underestimated when the radiation pressure from 
ionizing photons are neglected. In this work we have shown  
that the  BH masses for the RQ-BLS1s obtained from AD model fitting is similar to that 
obtained from virial method. Therefore the method of AD model fits
can be applied to find the BH masses of other AGN types.

This work clearly shows that NLS1s have BH masses and accretion rates
similar to BLS1s and the BH masses of $\gamma$-NLS1s in our sample are similar to blazars. The only major
difference that now  persists between $\gamma$-NLS1s  and 
FSRQs is related to their host galaxies. FSRQs are hosted by ellipticals
and the scarce observations available on  NLS1s point to ambiguity on their host 
galaxy type. NLS1s are preferentially hosted by spirals \citep{2018A&A...619A..69J}, 
but the hosts of some $\gamma$-NLS1s
such as FBQS J1644+2619 and PKS 1502+036 seem to be elliptical \citep{2017MNRAS.469L..11D,2018MNRAS.478L..66D}.
If future deep imaging observations do confirm that
$\gamma$-NLS1s are indeed hosted by spiral galaxies,
launching of relativistic jets in AGN is independent of their host galaxy type.
We do have reports of disk galaxies 
\citep{1998ApJ...495..227L,2011MNRAS.417L..36H,2015MNRAS.454.1556S} 
as well as RL-NLS1s (see \citealt{2018ApJ...869..173R} and references therein)
having large scale relativistic jets.

\section{Summary}
\begin{enumerate} 
\item We have estimated new BH masses using AD model fits and virial method for 
RQ-BLS1s, while for RQ-NLS1s and RL-NLS1s we have estimated new BH masses
using AD model fits.
\item From AD model fits, the mean estimated values of log($M_{BH}^{AD}/M_{\odot}$) for 
RQ-NLS1s  and RQ-BLS1s are 8.00$\pm$0.43 and 7.90$\pm$0.57, respectively. The
corresponding mean values obtained from  virial method are 7.07$\pm$0.38 and 8.01$\pm$0.48, respectively. 
\item For RL-NLS1s and RQ-NLS1s  we found that the BH masses estimated from AD model fits
are about an order of magnitude times larger than the BH masses obtained from virial method. However, for 
RQ-BLS1s, the BH masses obtained from AD model fits are in reasonable agreement to that 
obtained from virial method with a mean difference of log($M_{BH}^{AD}/M_{BH}^{VM}$) =  -0.11$\pm$0.64.

\item In our sample of  537 RL-NLS1s for which we were able to derive BH masses
from AD fitting, 
 9  are emitters of $\gamma$-rays. The mean values of log(M$_{BH}/M_{\odot}$) 
for these  9 sources from AD model fit and virial method are 8.15 $\pm$0.56 
and 7.28 $\pm$ 0.20, respectively. This indicates that $\gamma$-ray emitting NLS1s are
not low mass BH sources, instead have masses similar to 
blazars. 
\item 
NLS1s are not low mass BH and 
highly accreting sources as believed now, instead have BH masses and accretion rates similar to BLS1s.
\end{enumerate}

\acknowledgements
We thank the anonymous referee for his/her critical comments.

\bibliographystyle{apj}

\begin{thebibliography}{}
\expandafter\ifx\csname natexlab\endcsname\relax\def\natexlab#1{#1}\fi

\bibitem[{{Abdo} {et~al.}(2009){Abdo}, {Ackermann}, {Ajello}, {Axelsson},
  {Baldini}, {Ballet}, {Barbiellini}, {Bastieri}, {Battelino}, {Baughman},
  {Bechtol}, {Bellazzini}, {Bloom}, {Bonamente}, {Borgland}, {Bregeon}, {Brez},
  {Brigida}, {Bruel}, {Caliandro}, {Cameron}, {Caraveo}, {Casandjian},
  {Cavazzuti}, {Cecchi}, {Chekhtman}, {Cheung}, {Chiang}, {Ciprini}, {Claus},
  {Cohen-Tanugi}, {Collmar}, {Conrad}, {Costamante}, {Dermer}, {de Angelis},
  {de Palma}, {Digel}, {Silva}, {Drell}, {Dubois}, {Dumora}, {Farnier},
  {Favuzzi}, {Focke}, {Foschini}, {Frailis}, {Fuhrmann}, {Fukazawa}, {Funk},
  {Fusco}, {Gargano}, {Gehrels}, {Germani}, {Giebels}, {Giglietto}, {Giordano},
  {Giroletti}, {Glanzman}, {Grenier}, {Grondin}, {Grove}, {Guillemot},
  {Guiriec}, {Hanabata}, {Harding}, {Hartman}, {Hayashida}, {Hays}, {Hughes},
  {J{\'o}hannesson}, {Johnson}, {Johnson}, {Johnson}, {Kamae}, {Katagiri},
  {Kataoka}, {Kerr}, {Kn{\"o}dlseder}, {Kuehn}, {Kuss}, {Lande}, {Latronico},
  {Lemoine-Goumard}, {Longo}, {Loparco}, {Lott}, {Lovellette}, {Lubrano},
  {Madejski}, {Makeev}, {Max-Moerbeck}, {Mazziotta}, {McConville}, {McEnery},
  {Meurer}, {Michelson}, {Mitthumsiri}, {Mizuno}, {Monte}, {Monzani},
  {Morselli}, {Moskalenko}, {Murgia}, {Nolan}, {Norris}, {Nuss}, {Ohsugi},
  {Omodei}, {Orlando}, {Ormes}, {Paneque}, {Panetta}, {Parent}, {Pavlidou},
  {Pearson}, {Pepe}, {Pesce-Rollins}, {Piron}, {Porter}, {Rain{\`o}}, {Rando},
  {Razzano}, {Readhead}, {Reimer}, {Reimer}, {Reposeur}, {Richards}, {Ritz},
  {Rodriguez}, {Romani}, {Ryde}, {Sadrozinski}, {Sambruna}, {Sanchez},
  {Sander}, {Parkinson}, {Scargle}, {Schalk}, {Sgr{\`o}}, {Smith}, {Spandre},
  {Spinelli}, {Starck}, {Stevenson}, {Strickman}, {Suson}, {Tagliaferri},
  {Takahashi}, {Tanaka}, {Thayer}, {Thompson}, {Tibaldo}, {Tibolla}, {Torres},
  {Tosti}, {Tramacere}, {Uchiyama}, {Usher}, {Vilchez}, {Vitale}, {Waite},
  {Winer}, {Wood}, {Ylinen}, {Zensus}, {Ziegler}, {Fermi/LAT Collaboration},
  {Ghisellini}, {Maraschi}, {Tavecchio}, \& {Angelakis}}]{2009ApJ...699..976A}
{Abdo}, A.~A., {Ackermann}, M., {Ajello}, M., {et~al.} 2009, \apj, 699, 976

\bibitem[{{Baldi} {et~al.}(2016){Baldi}, {Capetti}, {Robinson}, {Laor}, \&
  {Behar}}]{2016MNRAS.458L..69B}
{Baldi}, R.~D., {Capetti}, A., {Robinson}, A., {Laor}, A., \& {Behar}, E. 2016,
  \mnras, 458, L69

\bibitem[{{Bentz} {et~al.}(2013){Bentz}, {Denney}, {Grier}, {Barth},
  {Peterson}, {Vestergaard}, {Bennert}, {Canalizo}, {De Rosa}, {Filippenko},
  {Gates}, {Greene}, {Li}, {Malkan}, {Pogge}, {Stern}, {Treu}, \&
  {Woo}}]{2013ApJ...767..149B}
{Bentz}, M.~C., {Denney}, K.~D., {Grier}, C.~J., {et~al.} 2013, \apj, 767, 149

\bibitem[{{Boller} {et~al.}(1996){Boller}, {Brandt}, \&
  {Fink}}]{1996A&A...305...53B}
{Boller}, T., {Brandt}, W.~N., \& {Fink}, H. 1996, \aap, 305, 53

\bibitem[{{Calderone} {et~al.}(2018){Calderone}, {D'Ammando}, \&
  {Sbarrato}}]{2018rnls.confE..44C}
{Calderone}, G., {D'Ammando}, F., \& {Sbarrato}, T. 2018, in Revisiting
  narrow-line Seyfert 1 galaxies and their place in the Universe. 9-13 April
  2018. Padova Botanical Garden, Italy. Online at <A
  href=``https://pos.sissa.it/cgi-bin/reader/conf.cgi?confid=328''>https://pos.sissa.it/cgi-bin/reader/conf.cgi?confid=328</A>,
  id.44, 44

\bibitem[{{Calderone} {et~al.}(2013){Calderone}, {Ghisellini}, {Colpi}, \&
  {Dotti}}]{2013MNRAS.431..210C}
{Calderone}, G., {Ghisellini}, G., {Colpi}, M., \& {Dotti}, M. 2013, \mnras,
  431, 210

\bibitem[{{Capellupo} {et~al.}(2015){Capellupo}, {Netzer}, {Lira},
  {Trakhtenbrot}, \& {Mej{\'{\i}}a-Restrepo}}]{2015MNRAS.446.3427C}
{Capellupo}, D.~M., {Netzer}, H., {Lira}, P., {Trakhtenbrot}, B., \&
  {Mej{\'{\i}}a-Restrepo}, J. 2015, \mnras, 446, 3427

\bibitem[{{Capellupo} {et~al.}(2016){Capellupo}, {Netzer}, {Lira},
  {Trakhtenbrot}, \& {Mej{\'{\i}}a-Restrepo}}]{2016MNRAS.460..212C}
---. 2016, \mnras, 460, 212

\bibitem[{{Cardelli} {et~al.}(1989){Cardelli}, {Clayton}, \&
  {Mathis}}]{1989ApJ...345..245C}
{Cardelli}, J.~A., {Clayton}, G.~C., \& {Mathis}, J.~S. 1989, \apj, 345, 245

\bibitem[{{Constantin} \& {Shields}(2003)}]{2003PASP..115..592C}
{Constantin}, A., \& {Shields}, J.~C. 2003, \pasp, 115, 592

\bibitem[{{D'Ammando} {et~al.}(2018){D'Ammando}, {Acosta-Pulido}, {Capetti},
  {Baldi}, {Orienti}, {Raiteri}, \& {Ramos Almeida}}]{2018MNRAS.478L..66D}
{D'Ammando}, F., {Acosta-Pulido}, J.~A., {Capetti}, A., {et~al.} 2018, \mnras,
  478, L66

\bibitem[{{D'Ammando} {et~al.}(2017){D'Ammando}, {Acosta-Pulido}, {Capetti},
  {Raiteri}, {Baldi}, {Orienti}, \& {Ramos Almeida}}]{2017MNRAS.469L..11D}
---. 2017, \mnras, 469, L11

\bibitem[{{D'Ammando} {et~al.}(2012){D'Ammando}, {Orienti}, {Finke}, {Raiteri},
  {Angelakis}, {Fuhrmann}, {Giroletti}, {Hovatta}, {Max-Moerbeck}, {Perkins},
  {Readhead}, {Richards}, {Stawarz}, \& {Donato}}]{2012MNRAS.426..317D}
{D'Ammando}, F., {Orienti}, M., {Finke}, J., {et~al.} 2012, \mnras, 426, 317

\bibitem[{{Decarli} {et~al.}(2008){Decarli}, {Dotti}, {Fontana}, \&
  {Haardt}}]{2008MNRAS.386L..15D}
{Decarli}, R., {Dotti}, M., {Fontana}, M., \& {Haardt}, F. 2008, \mnras, 386,
  L15

\bibitem[{{Foschini}(2011)}]{2011nlsg.confE..24F}
{Foschini}, L. 2011, in Narrow-Line Seyfert 1 Galaxies and their Place in the
  Universe, 24

\bibitem[{{Hota} {et~al.}(2011){Hota}, {Sirothia}, {Ohyama}, {Konar}, {Kim},
  {Rey}, {Saikia}, {Croston}, \& {Matsushita}}]{2011MNRAS.417L..36H}
{Hota}, A., {Sirothia}, S.~K., {Ohyama}, Y., {et~al.} 2011, \mnras, 417, L36

\bibitem[{{J{\"a}rvel{\"a}} {et~al.}(2018){J{\"a}rvel{\"a}},
  {L{\"a}hteenm{\"a}ki}, \& {Berton}}]{2018A&A...619A..69J}
{J{\"a}rvel{\"a}}, E., {L{\"a}hteenm{\"a}ki}, A., \& {Berton}, M. 2018, \aap,
  619, A69

\bibitem[{{Jiang} {et~al.}(2012){Jiang}, {Zhou}, {Ho}, {Yuan}, {Wang}, {Dong},
  {Jiang}, {Ji}, \& {Tian}}]{2012ApJ...759L..31J}
{Jiang}, N., {Zhou}, H.-Y., {Ho}, L.~C., {et~al.} 2012, \apjl, 759, L31

\bibitem[{{Klimek} {et~al.}(2004){Klimek}, {Gaskell}, \&
  {Hedrick}}]{2004ApJ...609...69K}
{Klimek}, E.~S., {Gaskell}, C.~M., \& {Hedrick}, C.~H. 2004, \apj, 609, 69

\bibitem[{{Komossa}(2007)}]{2007ASPC..373..719K}
{Komossa}, S. 2007, in Astronomical Society of the Pacific Conference Series,
  Vol. 373, The Central Engine of Active Galactic Nuclei, ed. L.~C. {Ho} \&
  J.-W. {Wang}, 719

\bibitem[{{Kshama} {et~al.}(2017){Kshama}, {Paliya}, \&
  {Stalin}}]{2017MNRAS.466.2679K}
{Kshama}, S.~K., {Paliya}, V.~S., \& {Stalin}, C.~S. 2017, \mnras, 466, 2679

\bibitem[{{L{\"a}hteenm{\"a}ki} {et~al.}(2017){L{\"a}hteenm{\"a}ki},
  {J{\"a}rvel{\"a}}, {Hovatta}, {Tornikoski}, {Harrison}, {L{\'o}pez-Caniego},
  {Max-Moerbeck}, {Mingaliev}, {Pearson}, {Ramakrishnan}, {Readhead}, {Reeves},
  {Richards}, {Sotnikova}, \& {Tammi}}]{2017A&A...603A.100L}
{L{\"a}hteenm{\"a}ki}, A., {J{\"a}rvel{\"a}}, E., {Hovatta}, T., {et~al.} 2017,
  \aap, 603, A100

\bibitem[{{Ledlow} {et~al.}(1998){Ledlow}, {Owen}, \&
  {Keel}}]{1998ApJ...495..227L}
{Ledlow}, M.~J., {Owen}, F.~N., \& {Keel}, W.~C. 1998, \apj, 495, 227

\bibitem[{{Leighly}(1999)}]{1999ApJS..125..317L}
{Leighly}, K.~M. 1999, \apjs, 125, 317

\bibitem[{{Marconi} {et~al.}(2008){Marconi}, {Axon}, {Maiolino}, {Nagao},
  {Pastorini}, {Pietrini}, {Robinson}, \& {Torricelli}}]{2008ApJ...678..693M}
{Marconi}, A., {Axon}, D.~J., {Maiolino}, R., {et~al.} 2008, \apj, 678, 693

\bibitem[{{Mej{\'{\i}}a-Restrepo} {et~al.}(2018){Mej{\'{\i}}a-Restrepo},
  {Lira}, {Netzer}, {Trakhtenbrot}, \& {Capellupo}}]{2018NatAs...2...63M}
{Mej{\'{\i}}a-Restrepo}, J.~E., {Lira}, P., {Netzer}, H., {Trakhtenbrot}, B.,
  \& {Capellupo}, D.~M. 2018, Nature Astronomy, 2, 63

\bibitem[{{Moran} {et~al.}(1996){Moran}, {Halpern}, \&
  {Helfand}}]{1996ApJS..106..341M}
{Moran}, E.~C., {Halpern}, J.~P., \& {Helfand}, D.~J. 1996, \apjs, 106, 341

\bibitem[{{Ojha} {et~al.}(2019){Ojha}, {Krishna}, \&
  {Chand}}]{2019MNRAS.483.3036O}
{Ojha}, V., {Krishna}, G., \& {Chand}, H. 2019, \mnras, 483, 3036

\bibitem[{{Osterbrock} \& {Pogge}(1985)}]{1985ApJ...297..166O}
{Osterbrock}, D.~E., \& {Pogge}, R.~W. 1985, \apj, 297, 166

\bibitem[{{Paliya} {et~al.}(2018){Paliya}, {Ajello}, {Rakshit}, {Mandal},
  {Stalin}, {Kaur}, \& {Hartmann}}]{2018ApJ...853L...2P}
{Paliya}, V.~S., {Ajello}, M., {Rakshit}, S., {et~al.} 2018, \apjl, 853, L2

\bibitem[{{Paliya} {et~al.}(2019){Paliya}, {Parker}, {Jiang}, {Fabian},
  {Brenneman}, {Ajello}, \& {Hartmann}}]{2019ApJ...872..169P}
{Paliya}, V.~S., {Parker}, M.~L., {Jiang}, J., {et~al.} 2019, \apj, 872, 169

\bibitem[{{Paliya} {et~al.}(2016){Paliya}, {Rajput}, {Stalin}, \&
  {Pandey}}]{2016ApJ...819..121P}
{Paliya}, V.~S., {Rajput}, B., {Stalin}, C.~S., \& {Pandey}, S.~B. 2016, \apj,
  819, 121

\bibitem[{{Paliya} {et~al.}(2014){Paliya}, {Sahayanathan}, {Parker}, {Fabian},
  {Stalin}, {Anjum}, \& {Pandey}}]{2014ApJ...789..143P}
{Paliya}, V.~S., {Sahayanathan}, S., {Parker}, M.~L., {et~al.} 2014, \apj, 789,
  143

\bibitem[{{Paliya} {et~al.}(2013){Paliya}, {Stalin}, {Kumar}, {Kumar}, {Bhatt},
  {Pandey}, \& {Yadav}}]{2013MNRAS.428.2450P}
{Paliya}, V.~S., {Stalin}, C.~S., {Kumar}, B., {et~al.} 2013, \mnras, 428, 2450

\bibitem[{{Rakshit} {et~al.}(2019){Rakshit}, {Johnson}, {Stalin}, {Gandhi}, \&
  {Hoenig}}]{2019MNRAS.483.2362R}
{Rakshit}, S., {Johnson}, A., {Stalin}, C.~S., {Gandhi}, P., \& {Hoenig}, S.
  2019, \mnras, 483, 2362

\bibitem[{{Rakshit} \& {Stalin}(2017)}]{2017ApJ...842...96R}
{Rakshit}, S., \& {Stalin}, C.~S. 2017, \apj, 842, 96

\bibitem[{{Rakshit} {et~al.}(2017){Rakshit}, {Stalin}, {Chand}, \&
  {Zhang}}]{2017ApJS..229...39R}
{Rakshit}, S., {Stalin}, C.~S., {Chand}, H., \& {Zhang}, X.-G. 2017, \apjs,
  229, 39

\bibitem[{{Rakshit} {et~al.}(2018){Rakshit}, {Stalin}, {Hota}, \&
  {Konar}}]{2018ApJ...869..173R}
{Rakshit}, S., {Stalin}, C.~S., {Hota}, A., \& {Konar}, C. 2018, \apj, 869, 173

\bibitem[{{Rani} {et~al.}(2017){Rani}, {Stalin}, \&
  {Rakshit}}]{2017MNRAS.466.3309R}
{Rani}, P., {Stalin}, C.~S., \& {Rakshit}, S. 2017, \mnras, 466, 3309

\bibitem[{{Rodriguez-Pascual} {et~al.}(1997){Rodriguez-Pascual}, {Mas-Hesse},
  \& {Santos-Lleo}}]{1997A&A...327...72R}
{Rodriguez-Pascual}, P.~M., {Mas-Hesse}, J.~M., \& {Santos-Lleo}, M. 1997,
  \aap, 327, 72

\bibitem[{{Ryan} {et~al.}(2007){Ryan}, {De Robertis}, {Virani}, {Laor}, \&
  {Dawson}}]{2007ApJ...654..799R}
{Ryan}, C.~J., {De Robertis}, M.~M., {Virani}, S., {Laor}, A., \& {Dawson},
  P.~C. 2007, \apj, 654, 799

\bibitem[{{Shakura} \& {Sunyaev}(1973)}]{1973A&A....24..337S}
{Shakura}, N.~I., \& {Sunyaev}, R.~A. 1973, \aap, 24, 337

\bibitem[{{Singh} {et~al.}(2015){Singh}, {Ishwara-Chandra}, {Sievers},
  {Wadadekar}, {Hilton}, \& {Beelen}}]{2015MNRAS.454.1556S}
{Singh}, V., {Ishwara-Chandra}, C.~H., {Sievers}, J., {et~al.} 2015, \mnras,
  454, 1556

\bibitem[{{Wang} {et~al.}(1996){Wang}, {Brinkmann}, \&
  {Bergeron}}]{1996A&A...309...81W}
{Wang}, T., {Brinkmann}, W., \& {Bergeron}, J. 1996, \aap, 309, 81

\bibitem[{{Williams} {et~al.}(2018){Williams}, {Gliozzi}, \&
  {Rudzinsky}}]{2018MNRAS.480...96W}
{Williams}, J.~K., {Gliozzi}, M., \& {Rudzinsky}, R.~V. 2018, \mnras, 480, 96

\bibitem[{{Yang} {et~al.}(2018){Yang}, {Yuan}, {Yao}, {Li}, {Zhang}, {Zhou},
  {Komossa}, \& {Liu}}]{2018arXiv180103963Y}
{Yang}, H., {Yuan}, W., {Yao}, S., {et~al.} 2018, ArXiv e-prints,
  arXiv:1801.03963

\bibitem[{{Yuan} {et~al.}(2008){Yuan}, {Zhou}, {Komossa}, {Dong}, {Wang}, {Lu},
  \& {Bai}}]{2008ApJ...685..801Y}
{Yuan}, W., {Zhou}, H.~Y., {Komossa}, S., {et~al.} 2008, \apj, 685, 801

\bibitem[{{Zhou} {et~al.}(2006){Zhou}, {Wang}, {Yuan}, {Lu}, {Dong}, {Wang}, \&
  {Lu}}]{2006ApJS..166..128Z}
{Zhou}, H., {Wang}, T., {Yuan}, W., {et~al.} 2006, \apjs, 166, 128

\end{thebibliography}

\end{document}